\documentclass[twocolumn,showpacs,showkeys,preprintnumbers,amsmath,amssymb]{revtex4}

\begin{document}
\draft{}

\bibliographystyle{try}

\topmargin 0.1cm
\hyphenation{pre-dicts}
\hyphenation{me-thods}
\hyphenation{seems}
\hyphenation{re-sults}
\hyphenation{li-mi-ted}
\hyphenation{pa-ra-me-ters}
\hyphenation{uni-ty}
\hyphenation{all}

 
 \newcounter{univ_counter}
 \setcounter{univ_counter} {0}

\addtocounter{univ_counter} {1} 
\edef\INFNFR{$^{\arabic{univ_counter}}$ } 

\addtocounter{univ_counter} {1} 
\edef\ROMA{$^{\arabic{univ_counter}}$ } 

\addtocounter{univ_counter} {1} 
\edef\ASU{$^{\arabic{univ_counter}}$ } 

\addtocounter{univ_counter} {1} 
\edef\SACLAY{$^{\arabic{univ_counter}}$ } 

\addtocounter{univ_counter} {1} 
\edef\UCLA{$^{\arabic{univ_counter}}$ } 

\addtocounter{univ_counter} {1} 
\edef\CMU{$^{\arabic{univ_counter}}$ } 

\addtocounter{univ_counter} {1} 
\edef\CUA{$^{\arabic{univ_counter}}$ } 

\addtocounter{univ_counter} {1} 
\edef\CNU{$^{\arabic{univ_counter}}$ } 

\addtocounter{univ_counter} {1} 
\edef\UCONN{$^{\arabic{univ_counter}}$ } 

\addtocounter{univ_counter} {1} 
\edef\DUKE{$^{\arabic{univ_counter}}$ } 

\addtocounter{univ_counter} {1} 
\edef\EDINBURGH{$^{\arabic{univ_counter}}$ } 

\addtocounter{univ_counter} {1} 
\edef\FIU{$^{\arabic{univ_counter}}$ } 

\addtocounter{univ_counter} {1} 
\edef\FSU{$^{\arabic{univ_counter}}$ } 

\addtocounter{univ_counter} {1} 
\edef\GWU{$^{\arabic{univ_counter}}$ } 

\addtocounter{univ_counter} {1} 
\edef\GLASGOW{$^{\arabic{univ_counter}}$ } 

\addtocounter{univ_counter} {1} 
\edef\INFNGE{$^{\arabic{univ_counter}}$ } 

\addtocounter{univ_counter} {1} 
\edef\ORSAY{$^{\arabic{univ_counter}}$ } 

\addtocounter{univ_counter} {1} 
\edef\ITEP{$^{\arabic{univ_counter}}$ }

\addtocounter{univ_counter} {1} 
\edef\JMU{$^{\arabic{univ_counter}}$ } 

\addtocounter{univ_counter} {1} 
\edef\KYUNGPOOK{$^{\arabic{univ_counter}}$ } 

\addtocounter{univ_counter} {1} 
\edef\MIT{$^{\arabic{univ_counter}}$ } 

\addtocounter{univ_counter} {1} 
\edef\UMASS{$^{\arabic{univ_counter}}$ } 

\addtocounter{univ_counter} {1} 
\edef\MSU{$^{\arabic{univ_counter}}$ } 

\addtocounter{univ_counter} {1} 
\edef\UNH{$^{\arabic{univ_counter}}$ } 

\addtocounter{univ_counter} {1} 
\edef\NSU{$^{\arabic{univ_counter}}$ } 

\addtocounter{univ_counter} {1} 
\edef\OHIOU{$^{\arabic{univ_counter}}$ } 

\addtocounter{univ_counter} {1} 
\edef\ODU{$^{\arabic{univ_counter}}$ } 

\addtocounter{univ_counter} {1} 
\edef\PITT{$^{\arabic{univ_counter}}$ } 

\addtocounter{univ_counter} {1} 
\edef\RPI{$^{\arabic{univ_counter}}$ } 

\addtocounter{univ_counter} {1} 
\edef\RICE{$^{\arabic{univ_counter}}$ } 

\addtocounter{univ_counter} {1} 
\edef\URICH{$^{\arabic{univ_counter}}$ } 

\addtocounter{univ_counter} {1} 
\edef\SCAROLINA{$^{\arabic{univ_counter}}$ } 

\addtocounter{univ_counter} {1} 
\edef\UTEP{$^{\arabic{univ_counter}}$ }

\addtocounter{univ_counter} {1} 
\edef\JLAB{$^{\arabic{univ_counter}}$ }  

\addtocounter{univ_counter} {1} 
\edef\UCONY{$^{\arabic{univ_counter}}$} 

\addtocounter{univ_counter} {1} 
\edef\VT{$^{\arabic{univ_counter}}$ } 

\addtocounter{univ_counter} {1} 
\edef\VIRGINIA{$^{\arabic{univ_counter}}$ } 

\addtocounter{univ_counter} {1} 
\edef\WM{$^{\arabic{univ_counter}}$ } 

\addtocounter{univ_counter} {1} 
\edef\YEREVAN{$^{\arabic{univ_counter}}$ } 

\title{{\large Onset of asymptotic scaling in deuteron photodisintegration}}
 \author{ 
P.~Rossi,\INFNFR\
M.~Mirazita,\INFNFR\
F.~Ronchetti,\INFNFR$^,$\ROMA\
E.~De~Sanctis,\INFNFR\ 
G.~Adams,\RPI\
P.~Ambrozewicz,\FIU\
E.~Anciant,\SACLAY\
M.~Anghinolfi,\INFNGE\
B.~Asavapibhop,\UMASS\
G.~Audit,\SACLAY
% ...
\\
(The CLAS collaboration)
} 
\address{\INFNFR Istituto Nazionale di Fisica Nucleare, Laboratori Nazionali di Frascati, PO 13, 00044 Frascati, Italy}
\address{\ROMA Universit\`{a} di ROMA III, 00146 Roma, Italy}
\address{\ASU Arizona State University, Tempe, Arizona 85287}
\address{\SACLAY CEA-Saclay, Service de Physique Nucl\'eaire, F91191 Gif-sur-Yvette, Cedex, France}
\address{\UCLA University of California at Los Angeles, Los Angeles, California  90095}
\address{\CMU Carnegie Mellon University, Pittsburgh, Pennsylvania 15213}
\address{\CUA Catholic University of America, Washington, D.C. 20064}
\address{\CNU Christopher Newport University, Newport News, Virginia 23606}
\address{\UCONN University of Connecticut, Storrs, Connecticut 06269}
\address{\DUKE Duke University, Durham, North Carolina 27708}
\address{\EDINBURGH Edinburgh University, Edinburgh EH9 3JZ, United Kingdom}
\address{\FIU Florida International University, Miami, Florida 33199}
\address{\FSU Florida State University, Tallahassee, Florida 32306}
\address{\GWU The George Washington University, Washington, DC 20052}
\address{\GLASGOW University of Glasgow, Glasgow G12 8QQ, United Kingdom}
\address{\INFNGE Istituto Nazionale di Fisica Nucleare, Sezione di Genova, 16146 Genova, Italy}
\address{\ORSAY Institut de Physique Nucleaire ORSAY, IN2P3 BP 1, 91406 Orsay, France}
\address{\ITEP Institute of Theoretical and Experimental Physics, Moscow, 117259, Russia}
\address{\JMU James Madison University, Harrisonburg, Virginia 22807}
\address{\KYUNGPOOK Kyungpook National University, Taegu 702-701, South Korea}
\address{\MIT Massachusetts Institute of Technology, Cambridge, Massachusetts  02139}
\address{\UMASS University of Massachusetts, Amherst, Massachusetts  01003}
\address{\MSU Moscow State University, 119899 Moscow, Russia}
\address{\UNH University of New Hampshire, Durham, New Hampshire 03824}
\address{\NSU Norfolk State University, Norfolk, Virginia 23504}
\address{\OHIOU Ohio University, Athens, Ohio 45701}
\address{\ODU Old Dominion University, Norfolk, Virginia 23529}
\address{\PITT University of Pittsburgh, Pittsburgh, Pennsylvania 15260}
\address{\RPI Rensselaer Polytechnic Institute, Troy, New York 12180}
\address{\RICE Rice University, Houston, Texas 77005}
\address{\URICH University of Richmond, Richmond, Virginia 23173}
\address{\SCAROLINA University of South Carolina, Columbia, South Carolina 29208}
\address{\UTEP University of Texas at El Paso, El Paso, Texas 79968}
\address{\JLAB Thomas Jefferson National Accelerator Facility, Newport News, Virginia 23606}
\address{\UCONY Union College, Schenectady, New York 12308}
\address{\VT Virginia Polytechnic Institute and State University, Blacksburg, Virginia 24061}
\address{\VIRGINIA University of Virginia, Charlottesville, Virginia 22901}
\address{\WM College of William and Mary, Williamsburg, Virginia 23187}
\address{\YEREVAN Yerevan Physics Institute, 375036 Yerevan, Armenia}
 
\date{\today}

\begin{abstract}
We investigate the transition from the nucleon-meson to quark-gluon description 
of the strong interaction using the photon energy dependence of the $d(\gamma,p)n$ differential 
cross section for photon energies above $0.5$~GeV and center-of-mass proton angles between $30^{\circ}$ 
and $150^{\circ}$. A possible signature for this transition is the onset of cross section $s^{-11}$ scaling 
with the total energy squared, $s$, at some proton transverse momentum, $P_T$. The results show that the 
scaling has been reached for proton transverse momentum above about 1.1~GeV/c. This may indicate that the quark-gluon regime is 
reached above this momentum.

\end{abstract}


\maketitle

\begin{figure*}[htb]
\vspace{18.cm}                                                                  
\includegraphics{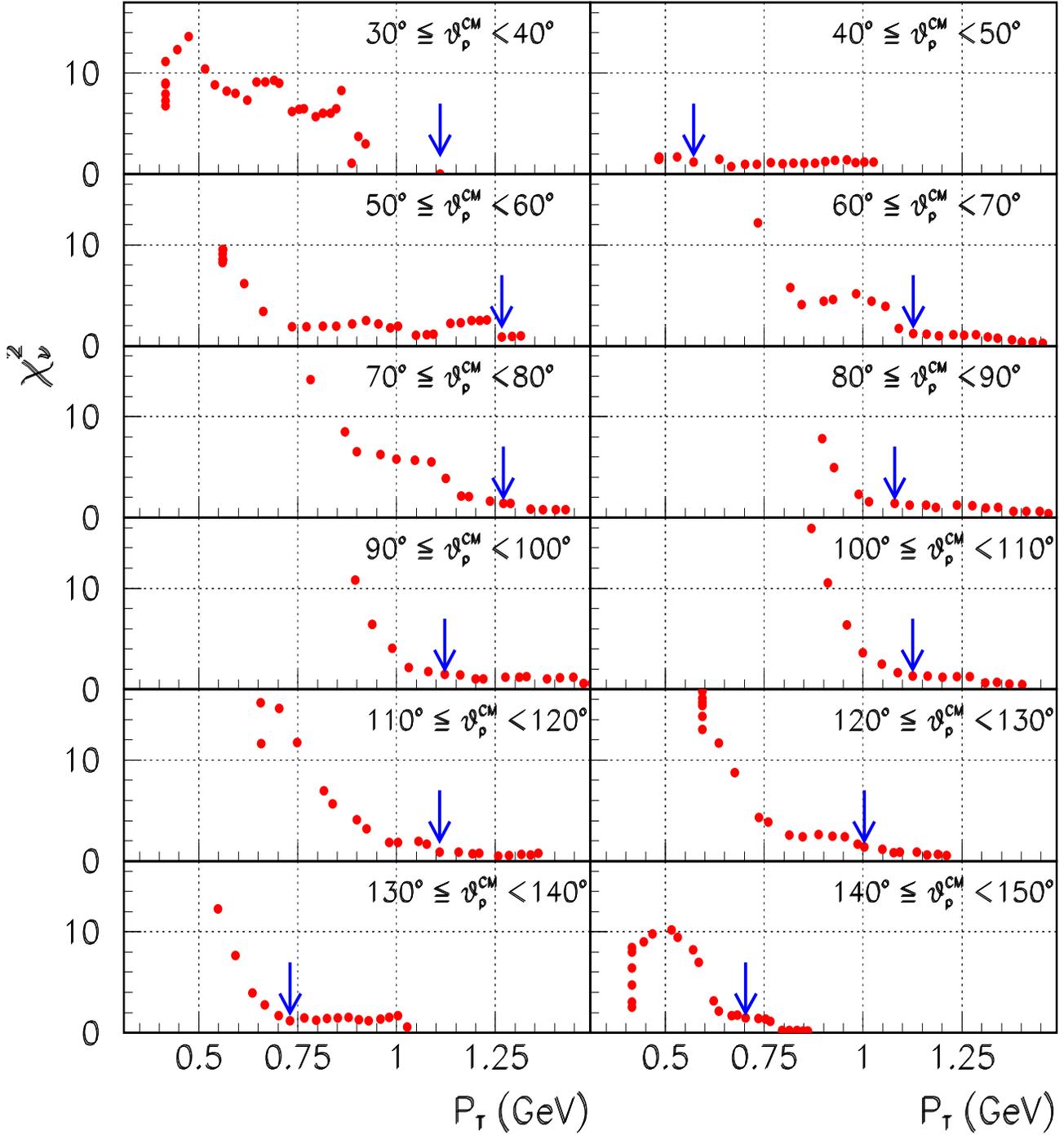}
\caption{\small Values of the reduced $\chi^2_\nu$ of the fits of the differential cross sections $d\sigma/dt$ in 
$\approx 1.2$~GeV $E_\gamma$ intervals with a power law $s^{-11}$, versus the related minimum proton transverse momentum 
$P_T$ for proton angles between $30^{\circ}$ and $150^{\circ}$. The vertical arrows indicate the transverse momentum 
thresholds for scaling.
}
\label{fig:enne1}
\end{figure*}

The interplay between the nucleonic and partonic pictures of the strong interaction 
represents one of the major issues in contemporary nuclear physics.
Although standard nuclear models are successful in describing the interactions between 
hadrons at large distances, and Quantum Chromodynamics (QCD) accounts well for the quark 
interactions at short distances, the physics connecting the two regimes 
remains unclear. 
In fact, the classical nucleonic description must break down once the probing distances 
become comparable to those separating the quarks. 
The challenge is to study this transition region by looking for the onset of some 
experimentally accessible phenomena  naturally predicted by perturbative QCD (pQCD).
The simplest is the constituent counting rule (CCR) for high energy exclusive 
reactions~\cite{Brodsky,Matveev}, in which $d\sigma/dt \propto s^{-n+2}$, with $n$ the 
total number of pointlike particles and gauge fields in the initial plus final states. 
Here $s$ and $t$ are the invariant Mandelstam variables for the 
total energy squared and the four-momentum transfer squared, respectively.\\
\indent
Deuteron photodisintegration is especially suited for this study, because a relatively large amount of 
momentum is transferred to the nucleons for a relatively low incident photon energy \cite{holt,gg}.
This reaction received renewed interest after an apparent onset of the expected 
asymptotic $s^{-11}$ scaling of the cross-section was observed at SLAC~\cite{SLACNE81,SLACNE82}
at center-of-mass proton scattering angle $\vartheta_p^{\rm{CM}}= 90^{\circ}$ and at 
about $E_\gamma$ =1~GeV photon energy. (For this reaction $n=13$, as there is one photon and 6+6=12 quarks).
Following this initial result, additional measurements were performed at SLAC~\cite{SLACNE17} and more 
recently at Thomas Jefferson National Accelerator Facility (TJNAF)~\cite{Bochna,Schulte,Gilman1,Gilman2,Mirazita} 
using different experimental techniques.
These data cover only a few proton angles. They show that a transition to QCD scaling 
seems to exist, but its boundaries are not well-defined. Scaling seems to be confirmed for center-of-mass 
proton angles $\vartheta_p^{\rm{CM}} = 69^{\circ}$ and $89^{\circ}$ \cite{Bochna} already at 
$E_{\gamma} = 1$~GeV photon energies, while at the forward angles  $\vartheta_p^{\rm{CM}} = 52^{\circ}$ 
and $36^{\circ}$, the cross section falls off more slowly than $s^{-11}$ until about 3 and 4 GeV beam 
energies, respectively~\cite{Schulte}.\\ 
\indent
The recent, extensive cross section data obtained at the TJNAF by CLAS experiment E93-017 between 0.5 and 
3.0 GeV with nearly complete proton angular coverage offer the unique opportunity for a detailed study of 
the energy dependence of the $d(\gamma,p)n$ differential cross section at fixed angles. 
A detailed description of the measurement and results has been reported in a separate paper~\cite{Mirazita}. 
Here we only point out that these data are consistent with previous measurements, and systematically cover the 
whole photon energy regime of interest.\\
\indent
In this paper we present the results of a detailed study of the behavior of $d\sigma/dt$ at fixed proton 
angle, $\vartheta_p^{CM}$, made to check the CCR $s^{-11}$ prediction as a function of the center-of-mass 
proton transverse momentum
\begin{equation}
P_T = \sqrt{\frac{1}{2}E_{\gamma}M_d\sin^2(\vartheta_p^{CM})},
\label{eq:pt}
\end{equation}
in which $M_d$ is the deuteron mass. $P_T$ is the correct kinematical 
variable for determining the onset of scaling \cite{brohi,carl}.\\ 
\indent
Differential cross sections $d\sigma/dt$ obtained above 0.5~GeV for fixed $\vartheta_p^{\rm{CM}}$ from all existing
high-energy  $\gamma d \rightarrow pn$ 
experiments~\cite{Bochna,Schulte,Gilman2,mainz,SLACNE81,SLACNE82,SLACNE17,Mirazita} 
have been grouped in $10^\circ$ wide bins and then fit to a power law $s^{-11}$ (one free parameter).
Table~\ref{tab:1} gives the photon energies and the proton angles where the differential cross sections 
have been measured by the experiments.
\begin{table}[tbh]
\caption{Photon energies and center-of-mass proton angles of the $\gamma d \rightarrow p n $ experiments 
whose data are used in the present work. }
\label{tab:1}
\begin{center}
\begin{tabular}{lllllll}
\hline\noalign{\smallskip}
        Exp.    & & & $E_{\gamma}$ (GeV)                   & & &$\vartheta_p^{CM}$(deg)\\
\noalign{\smallskip}\hline\noalign{\smallskip}
\cite{mainz}    & & & $0.5 - 0.78$                         & & & $40-160$             \\
\cite{Mirazita} & & & $0.5 - 3.0 $                         & & & $ 10 - 160 $         \\
\cite{SLACNE81} & & & 0.8, 1.1, 1.3, 1.6                   & & & 90                   \\
\cite{SLACNE82} & & & 0.8, 1.0, 1.2                        & & & 52, 66, 78, 90, 113, \\
                & & &                                      & & & 126, 142              \\
                & & & 1.4, 1.6, 1.8                        & & & 90, 113, 142         \\
\cite{Bochna}   & & & 0.8, 1.5, 2.4, 3.2, 4.0              & & & 36, 52, 69, 89       \\
\cite{SLACNE17} & & & 1.5, 1.9, 2.3, 2.7                   & & & 37, 53, 89           \\
\cite{Gilman2}  & & & 1.6, 1.9, 2.4                        & & & 30, 36, 52, 70, 90,  \\
                & & &                                      & & & 110, 127, 142          \\
\cite{Schulte}  & & & 5.0, 5.5                             & & & 37, 53, 70           \\
\noalign{\smallskip}\hline
\end{tabular}
\end{center}
\end{table}
Data were considered without any renormalization to each other, and with their statistical and systematic errors 
added in quadrature. In order to determine whether, and at which proton transverse momentum threshold, $P_T^{th}$, 
data start to follow the power law $s^{-11}$, fits were performed for partial samples of the data over about 
$1.2$~GeV wide windows in $E_\gamma$. 
These energy windows correspond to $P_T$ intervals of $200-400$~MeV/c, depending on the photon energy and the proton angle. 
(For fixed $\vartheta_p^{CM}$, $P_T$, $E_\gamma$ and $s$ are directly related, and each variable can be used 
interchangeably). The window in $E_\gamma$ was then shifted by 100 MeV, and another fit made. The process
was repeated up to the highest $E_\gamma$  window.\\
\indent
Figure~\ref{fig:enne1} shows the reduced $\chi^2_\nu$ values of the fits versus the related transverse proton momentum $P_T$ 
corresponding to the lower $E_\gamma$-value of each interval for $\vartheta_p^{CM}$ between $30^\circ$ and 
$150^\circ$. 
We limited the study to these angles because the data at more forward and backward angles lack the statistics 
for fits over a significant $P_T$ interval. These results are not 
changed significantly by the size of the $E_\gamma$ window, which if too large makes the fit insensitive to 
deviations from $s^{-11}$ at low $s$, and if too small makes it not reliable.

\begin{figure*}[htb]
\vspace{18.cm}                                                                  
\includegraphics{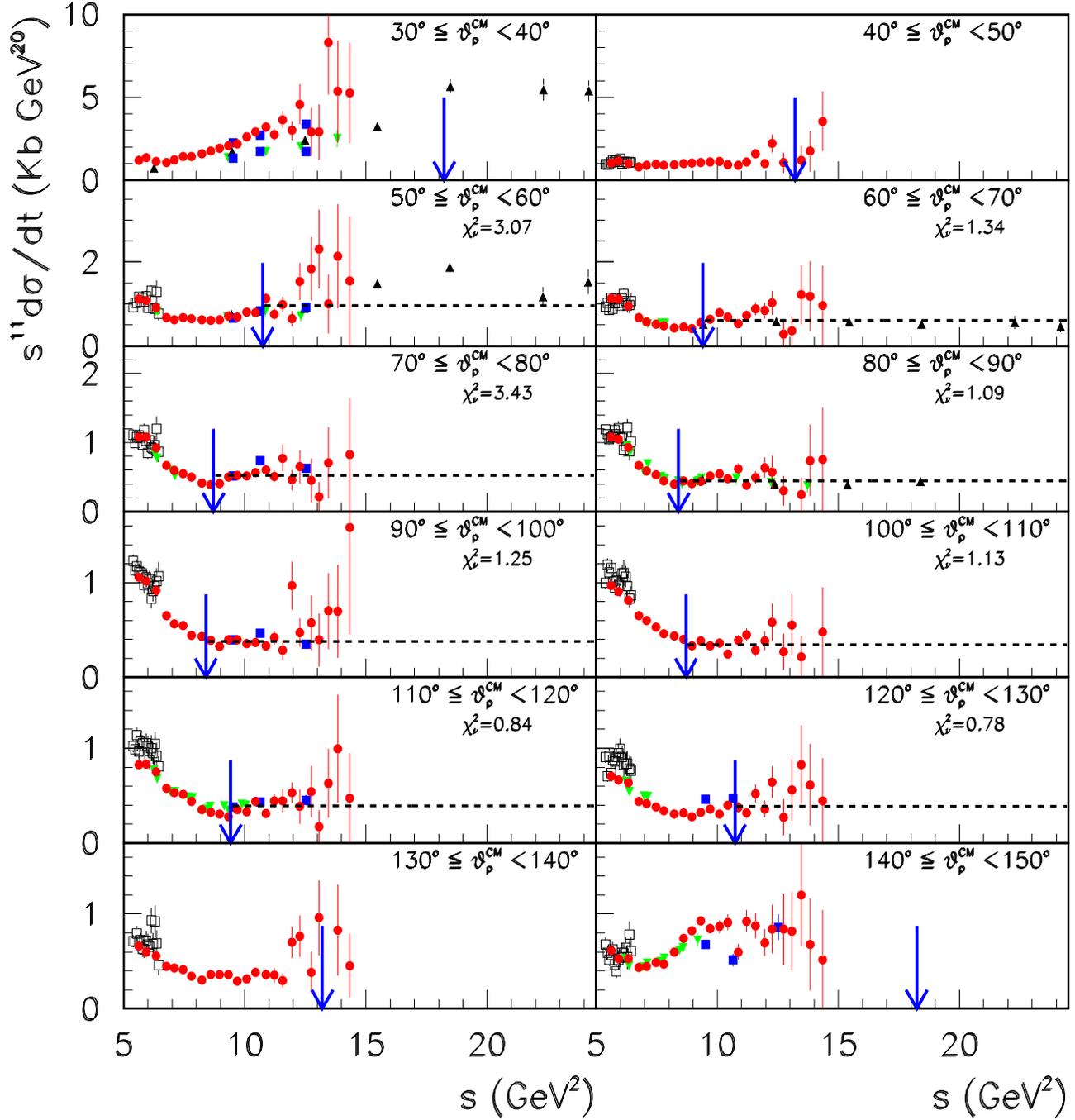}
\caption{\small  
Deuteron photodisintegration cross section, $s^{11}d\sigma/dt$, as a function of $s$ for the given proton scattering 
angles. Dashed lines are the fits of the data to $s^{-11}$ for $P_T \ge 1.1$~GeV/c. The vertical arrows indicate the 
$s$ value corresponding to $P_T=1.1$~GeV/c. Fits are not shown for $\vartheta_p^{CM} =35^\circ, 45^\circ$, $135^\circ$, 
and $145^\circ$ where there are not enough data above 1.1~GeV/c. Also shown in each panel is the  $\chi^2_\nu$ value of 
the fit. Data are from CLAS~\cite{Mirazita} (full/red circles),
Mainz~\cite{mainz} (open/black squares), SLAC~\cite{SLACNE17,SLACNE81,SLACNE82} 
(full-down/green triangles), JLab Hall~A~\cite{Gilman2} (full/blue squares) and 
Hall~C~\cite{Bochna,Schulte} (full-up/black triangles). 
}
\label{fig:fitfin}
\end{figure*}

\indent
Apart from $45^\circ$, where the $\chi^2_\nu$ is approximately constant around unity over the full $P_T$ range, at all 
other angles, the $\chi^2_\nu$ decreases from values $\geq 10$ at low $P_T$ towards unity at some $P_T^{th}$, and then 
remains approximately flat up to the highest $P_T$. 
Clearly, $P_T^{th}$ is the value above which the cross sections have a reliable $s^{-11}$ dependence.\\
\indent
The $10^\circ$ wide angular bins, the 100 MeV wide shifts among the $E_\gamma$ windows over which the fits are done, and the slow variation in $\chi^2_\nu$ 
do not allow the extraction of a precise $P_T^{th}$ for this transition. Nevertheless, one can evaluate an approximate 
value of $P_T^{th}$ by using a statistical criterion.
Specifically, for each angle a $\chi^2_\nu(90\%)$ ($\approx$~1.4--1.6, depending on the number of data points) has been fixed, 
corresponding to a $90\%$ confidence level for the fit; the transverse momentum threshold for scaling, $P_T^{th}$, has been 
chosen where $\chi^2_\nu$ of the fit becomes less or equal to the value $\chi^2_\nu(90\%)$.  The values of $P_T^{th}$ are 
shown by the vertical arrows in Fig.~\ref{fig:enne1}. They range between 1.00 and 1.27~GeV/c (average value 1.13~GeV/c)
at $35^\circ$ and in the angular bins between $50^\circ$ and $130^\circ$, and are about 0.6-0.7~GeV/c, at $45^\circ$,
$135^\circ$ and $145^\circ$. The uncertainties on $P_T$ values, estimated by changing the confidence level of the fits by 
$\pm 5\%$, are up to 80 MeV/c. However, this would seem to be an underestimate of the uncertainty given a visual inspection 
of Fig.~\ref{fig:enne1}. In particular, the uncertainty on $P_T$ is larger for the extreme angles ($35^\circ$, $45^\circ$, $135^\circ$ and $145^\circ$), where the derivative of $\sin(\vartheta_p^{CM})$ over the $10^\circ$ width of the angular bin is larger. (From Eq.~\ref{eq:pt} it results that $P_T$ is proportional to  $\vartheta_p^{CM}$). Overall, we believe that a reasonable uncertainty is larger than 100~MeV/c.\\
\indent
Then, to further check the consistency of data to the CCR prediction, we have fit all cross section data at fixed proton 
angle between $55^\circ$ and $125^\circ$ and $P_T \ge 1.1$~GeV/c to $s^{-11}$.
 We limited the fit to these angles, because 
at $\vartheta_p^{CM} = 35^\circ, 45^\circ$, $135^\circ$, and $145^\circ$ there are not enough data above $P_T$= 1.1 GeV/c 
to make a reliable fit.
These fits are shown in Fig.~\ref{fig:fitfin} together with the data from all the high-energy 
$\gamma d \rightarrow p n $ experiments \cite{Bochna,Schulte,Gilman2,mainz,SLACNE81,SLACNE82,SLACNE17,Mirazita} used in 
this study. 
For a sake of clearness, data have been multiplied by $s^{11}$. The $\chi^2_\nu$ of the fits are given in the plots. 
The vertical arrows indicate the $s$ value corresponding to $P_T=1.1$~GeV/c. It is worth noticing that for 
$\vartheta_p^{CM} = 35^\circ$ the last three points show a clear flat behaviour well consistent with an $s^{-11}$-dependence, 
as it is proven by the very low value $\chi_\nu^2$ = 0.03 of the last $P_T$ bin (1.10-1.30 GeV/c) in the first panel of 
Fig.~\ref{fig:enne1}.\\
\indent
For all but two of the fits, $\chi^2_\nu \leq 1.34$.  
At $55^\circ$, and in particular at $75^\circ$ the worse $\chi^2_\nu$ could be due to discrepancies in the absolute values 
of data from various experiments. As an example, the fit for $75^\circ$ with the data sets \cite{Gilman2,Mirazita}
renormalized to each other gives a $\chi^2_\nu$ =2.51.
This shows that the $s^{-11}$ dependence of the cross section is established for $P_T \geq$ 1.1~GeV/c. This is a 
necessary condition for the transition to the QCD scaling. Then, one might argue that the quark-gluon regime is reached 
for the proton transverse momenta above about 1.1~GeV/c.\\
\indent
In conclusion, the new, nearly complete angular distributions of two-body deuteron photodisintegration---obtained 
by CLAS at TJNAF for photon energies between 0.5 and 3.0 GeV---have been used, together with all previous 
data, for a detailed study of the power law $s$-dependence of the differential cross section.
The results show that the $s^{-11}$ scaling has been reached 
for proton transverse momentum above about 1.1~GeV/c. This may indicate that the quark-gluon regime is 
reached above this momentum.\\
\indent
This work was supported in part by the Italian Istituto Nazionale di Fisica Nucleare, 
the French Centre National de la Recherche Scientifique and the Commissariat \`{a} l'Energie 
Atomique, the U.S. Department of Energy and the National Science Foundation,
and the Korea Science and Engineering Foundation.

\end{document}